\documentclass[aps,prl,10pt,twocolumn,showpacs,superscriptaddress]{revtex4-1}
\usepackage{amssymb,graphicx,color,amsmath}
\usepackage[utf8x]{inputenc}
\usepackage{amsmath}
\usepackage{array}
\definecolor{lightblue}{rgb}{0.17,0.39,1}
\usepackage[bookmarks, colorlinks=true, breaklinks]{hyperref}
\hypersetup{linkcolor=lightblue, citecolor=lightblue, filecolor=black, urlcolor=lightblue}	

\begin{document}

\title {Quantum Critical Scaling of Specific Heat in a Quasicrystal}

\author{A.~Khansili}
\email[Email: ]{akash.khansili@fysik.su.se}
\affiliation{Department of Physics, Stockholm University, SE-10691 Stockholm, Sweden}

\author{Y.-C.~Huang}
\affiliation{Department of Chemistry, Uppsala University, 75121 Uppsala, Sweden}

\author{U. Häussermann}
\affiliation{Department of Materials and Environmental Chemistry, Stockholm University, S-10691 Stockholm, Sweden}

\author{C. Pay~Gomez}
\affiliation{Department of Chemistry, Uppsala University, 75121 Uppsala, Sweden}

\author{A.~Rydh}
\email[Email: ]{andreas.rydh@fysik.su.se}
\affiliation{Department of Physics, Stockholm University, SE-10691 Stockholm, Sweden}

\begin{abstract}
In strongly correlated systems, interactions give rise to critical fluctuations surrounding the quantum critical point (QCP) of a quantum phase transition. Quasicrystals allow the study of quantum critical phenomena in aperiodic systems with frustrated magnetic interactions. Here, we study the magnetic field and temperature scaling of the low-temperature specific heat for the quantum critical Yb-Au-Al quasicrystal. We devise a scaling function that encapsulates the limiting behaviors as well as the area where the system goes from a temperature-limited to a field-limited quantum critical region, where magnetic field acts as a cutoff for critical fluctuations. The zero-field electronic specific heat is described by a power-law divergence, ${C_{el}/T \propto T^{-0.54}}$, aligning with previously observed ac-susceptibility \cite{deguchi2012quantum} and specific heat measurements \cite{Watanuki2012}. The field dependence of the electronic specific heat at high magnetic fields shows a similar power-law ${C_{el}/T \propto B^{-0.50}}$. In the zero-field and low-field region, we observe two small but distinct anomalies in the specific heat, located at 0.7\,K and 2.1\,K.
\end{abstract}

\date{\today}
\maketitle

\section{Introduction}
Quantum critical systems cannot be understood in a simple Landau Fermi liquid (LFL) picture of conventional metals \cite{Stewart2001, Lohneysen2007, Gegenwart2008, Amusia2014}. Instead, the systems show non-Fermi liquid (NFL) behavior at finite temperatures above a quantum critical point (QCP) \cite{Schofield1999, Stewart2001}. This quantum critical region is observed in diverse systems, including high-$T_c$ superconductors \cite{Armitage2010, Abrahams2011, Varma2020}, heavy-fermion metals \cite{Amusia2014, Varma2020, Khansili2024quantum}, organic quantum materials \cite{Stone2006, Dressel2011, Furukawa2015}, as well as in the Yb-Au-Al quasicrystal \cite{deguchi2012quantum, Watanuki2012} which is studied here.

Quasicrystals \cite{shechtman1984metallic} allow the study of functional properties in aperiodic systems \cite{Wessel2003, Goldman2013, Goldman2014, Thiem2015, Kamiya2018, Tamura2021, Sato2022, Takeuchi2023, Lopez2023, Terashima2024}. The inclusion of rare-earth elements in these systems can induce nontrivial properties due to the localized $4\mathrm{f}$ electrons \cite{Charrier1997, Thiem2015, Tamura2021, Sato2022}. The interaction of conduction electrons with the localized $\mathrm{f}$-electrons opens for strong correlations within these systems, classifying them as heavy-fermion materials \cite{Stewart2001, Lohneysen2007, Gegenwart2008, Amusia2014, Sato2022}. Such strongly correlated electron systems can be tuned using a tuning parameter, often chemical doping or pressure, so that the systems are brought close to the QCP \cite{Pham2006, Lohneysen2007, Gegenwart2008, Amusia2014}.

 The electronic specific heat for a quantum critical system across temperatures and magnetic fields is expected to be understood in terms of quantum critical scaling. The critical behavior of the electronic specific heat in zero field is often described by a critical exponent $\alpha$, $C_{\mathrm{el}}/T \propto T^{-\alpha}$ \cite{Belitz2005, Lohneysen2007, Gegenwart2008, Amusia2014}. Power-law behavior has been observed for Yb-Au-Al in low-temperature ac-susceptibility measurements where $\chi_\mathrm{ac} \propto T^{-0.51}$ \cite{deguchi2012quantum} and previous specific heat measurements $C_{\mathrm{el}}/T \propto T^{-0.66}$ \cite{Watanuki2012}. 
 Here, we study the effect of magnetic field on the low-temperature specific heat of Yb-Au-Al, and describe the corresponding scaling behavior of its quantum criticality.

\section{Experimental Details}

Yb-Au-Al quasicrystals were synthesized using arc-melting, aiming for a nominal composition of Yb$_{17}$Au$_{49}$Al$_{34}$, following the strategy outlined in \cite{Ishimasa2011}. Metal mixtures composed of Yb (99.99\%), Au (99.99\%), and Al (99.999\%) granules were arc-melted. A 1\,g homogeneous alloy ingot was produced after repeating the arc-melting procedure several times. The mass loss was limited to within 2 weight percent. The alloy ingot was characterized by powder X-ray diffraction and energy dispersive spectroscopy using a scanning electron microscope. The characterized composition of the ingot was Yb$_{15}$Au$_{53}$Al$_{32}$, with the quasilattice parameter indexed at $a_{\mathrm{6D}} = 7.449$\,{\AA}. Elemental mapping showed a homogeneously distributed composition and the indexed powder pattern with the corresponding six indices of the Yb-Au-Al quasicrystal was consistent with previous reports \cite{deguchi2012quantum}.

\begin{figure*}[t!!]
\centering
\includegraphics[width=1\textwidth]{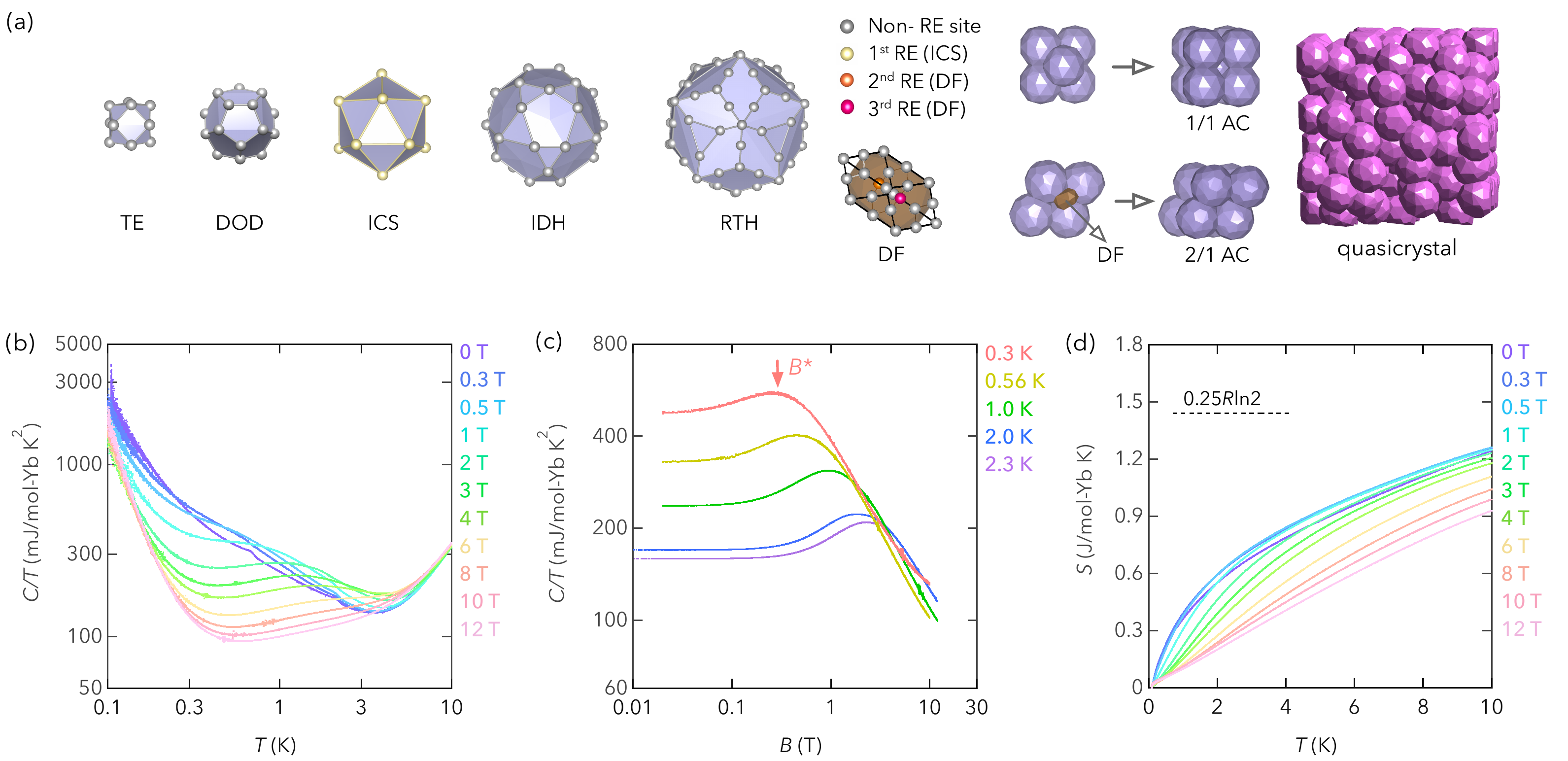}
\caption{Measured specific heat as a function of temperature and magnetic field.	 %
(a) The polyhedron cluster shells that make the Tsai cluster which is arranged in a periodic or aperiodic fashion to build the approximant crystals (AC) and quasicrystals. The clusters contain tetrahedron (TE), dodecahedron (DOD), icosahedron (ICS), icosidodecahedron (IDH), rhombic triacontahedron (RTH), and double Friauf (DF). 
(b) Temperature dependence of specific heat at different magnetic fields between 0.1-10\,K. The specific heat at high fields below 0.3\,K comes from nuclear contribution. The effect of nuclear contribution on the total specific heat above 0.3\,K is very small and only becomes significant at the highest fields.
(c) Magnetic field dependence of specific heat at several temperatures. Specific heat shows a temperature dependent maximum indicated by $B^\ast$ in the field dependence for 0.3\,K.
(d) Entropy as integrated $C_\mathrm{el}/T$ after subtracting phonon and nuclear contributions from the total $C/T$ in panel a.}
\label{fig:1}
\end{figure*}

Specific heat was measured using a nanocalorimeter \cite{tagliati2012differential, willa2017nanocalorimeter} operating in a frequency-adjusted ac steady-state mode \cite{tagliati2012differential}. The sample was $\sim6.2$\,nmol in size and attached to the calorimeter using Apiezon-N grease.
To verify that the ac calorimetry method captures all heat capacity components at low temperatures, we studied the frequency dependence at low temperatures using thermal impedance spectroscopy (TISP) \cite{Khansili2023}. The TISP measurements were performed on the same sample using a frequency range of $0.1$\,Hz - $3$\,kHz. The nuclear spins subsystem is connected to the electronic system via the nuclear spin-lattice relaxation time $T_1$. We found that the nuclear spin-lattice relaxation time ($T_1$) in Yb-Au-Al is fast enough, i.e., $T_1 < 1$\,ms, to always completely include the full nuclear contribution in the specific heat obtained from ac calorimetry.

\section{Results \& Discussion}
The quasicrystal and related approximants (1/1 AC and 2/1 AC) are the aperiodic and periodic arrangement of the Tsai cluster \cite{tsai2000stable, Takakura2007} comprised of the polyhedrons described in Fig.~\ref{fig:1}. The rare earth Yb atoms are located in the icosahedron (ICS) and double Friauf (DF) shells. The magnetic state of Yb$^{3+}$ is $J=7/2$ in a 4f$^{13}$ configuration. The non-cubic crystal field splits the state into four Kramers doublets. Magnetic susceptibility measurements \cite{Watanuki2012} indicate that the crystal field splitting becomes important below about $50$\,K, below which the magnetic behavior of the ion will be dominated by a single Kramers doublet. Such effective spin-1/2 ions are known to be important for the emergence of magnetic quantum criticality in heavy fermion systems \cite{Coleman2010, Coleman2010quantum, Gegenwart2015}.

Figure~\ref{fig:1}b and \ref{fig:1}c show the temperature and magnetic field dependence of the specific heat as $C/T$ of the Yb-Au-Al quasicrystal.
Fig.~\ref{fig:1}b shows the temperature dependence for several different magnetic fields in the range 0.1-10\,K. The phonon contribution dominates the specific heat at high temperatures ($T> 5$\,K). The low-temperature upturn in the specific heat at high magnetic field is due to a nuclear contribution. In zero-field, at temperatures below 0.3\,K, the nuclear specific heat starts to become comparable to the electronic specific heat. The nuclear contribution comes from Yb, Au, and Al where $^{173}$Yb has nuclear spin $I = 5/2$, $^{197}$Au with nuclear spin $I = 3/2$, and $^{27}$Al with $I = 5/2$ \cite{Stone2016}. All these elements have a non-zero electric quadrupole moment \cite{Stone2016}. Among these, $^{173}$Yb has the largest quadrupole moment of $Q = +2.80$\,barn \cite{Stone2016} that gives the low-temperature nuclear specific heat at zero field.

The zero-field electronic specific heat follows a power-law divergence as the temperature is decreased below 5\,K. As the magnetic field is applied, the specific heat starts to show suppression from this diverging behavior. At the highest field of $12$\,T, the electronic specific heat is suppressed to a value of $\sim 100$\,mJ/mol-Yb\,K$^2$ at 0.3\,K.

The magnetic field dependence for several temperatures is shown in Fig.~\ref{fig:1}c, illustrating that $C/T$ becomes constant below a temperature-dependent crossover field $B^\ast$, shown as an arrow for 0.3\,K, at which $C/T$ shows a maximum. The $C/T$ is suppressed above this crossover field up to the highest applied magnetic fields at all studied temperatures. This effect of field and temperature to cut off the enhancement of the electronic density of states is a prominent feature of quantum critical systems \cite{Lohneysen2001, Lohneysen2007, Amusia2014, Khansili2023multi}. Similar behavior is seen in the magnetic susceptibility of this system \cite{deguchi2012quantum} to the lowest measured temperatures, establishing that the system is intrinsically close to a quantum critical point.

The nuclear specific heat is dominated by the zero-field quadrupole contribution at all studied fields and can be considered field independent in these field ranges. As seen from Fig.~\ref{fig:1}b and \ref{fig:1}c, there is, thus, approximately no increase of $C/T$ with field at high magnetic fields.

Figure~\ref{fig:1}d shows the entropy integrated from $C_{el}/T$ after subtracting nuclear and phonon contributions. The entropy at 10\,K is clearly below the value $R\ln 2$ expected from a non-interacting Kramers doublet. Electronic correlations are thus important and have significantly affected the behavior of the Kramers doublet. We take this as support for describing the system as quantum critical.

At $T = 0$, a system at the quantum critical point is effectively described by a ($d_\mathrm{eff} = d+z$) dimensional classical system, where $d$ is the spatial dimension and $z$ is the dynamic exponent related to the additional imaginary time dimension \cite{Hertz1976, Millis1993}. At the quantum critical phase transition ($T = 0$) the associated correlation length $\xi$ and correlation time $\xi_t$ diverge \cite{Hertz1976, Millis1993, Lohneysen2007}, 
\begin{align}\label{Eq:CorrlationLength}
    \xi = |r|^{-\nu}, \hspace{1cm} \xi_t = \xi^z = |r|^{-\nu z}
\end{align}
where $r$ is the control parameter for quantum criticality (such as composition or pressure) describing the distance from the quantum critical point, $\nu$ is the critical exponent related to the correlation length, and $z$ is the dynamic exponent \cite{Hertz1976, Millis1993}.  

A finite temperature ($T>0$) acts as a regulator for the length of the time dimension as ${L_t = \hbar/k_B T}$ \cite{Hertz1976, Millis1993}.
There is, thus, an effective correlation time $\xi_{t,\mathrm{eff}}$ that can be expressed as 
\begin{equation}\label{Eq:XiEffective}
\xi_{t, \mathrm{eff}} \simeq \min\{\xi_t, L_t \}
\end{equation}
over the entire range of temperatures.
When $r=0$ or for $r \ne 0$ at high temperature, i.e. where $L_t \ll \xi_t$, the effective correlation time is cut off by $L_t$ (the so called quantum critical region). Even if a system is not located exactly at the QCP, the system behavior will be experimentally indistinguishable from a quantum critical system at the QCP as long as $\xi_{t, \mathrm{eff}}$ is given by $L_t$.

\begin{figure*}[t!!]
\centering
\includegraphics[width=1\textwidth]{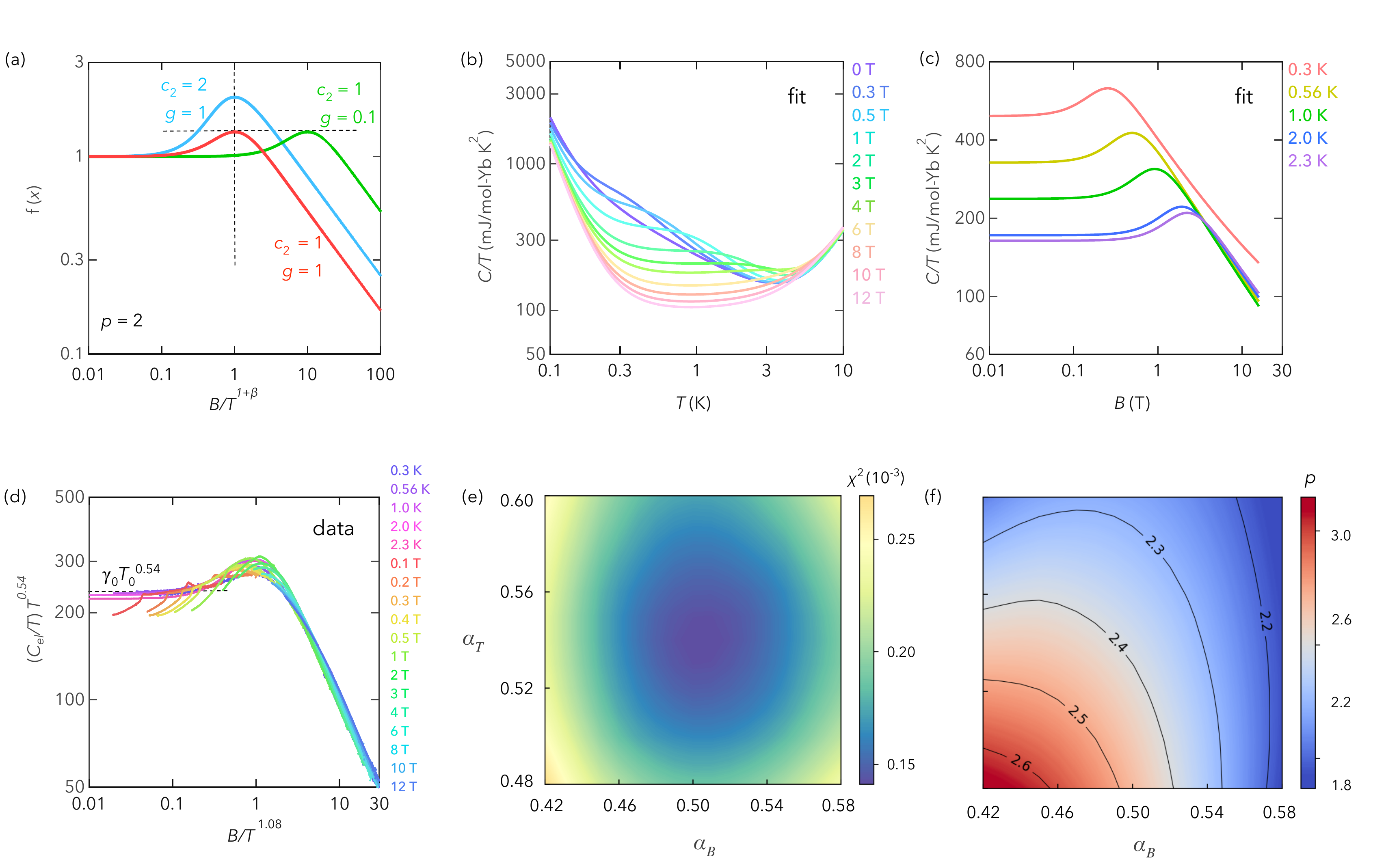} 
\caption{Analysis of scaling behavior.	
(a) Scaling function $f(x)$ given in Eq.~(\ref{Eq:ScalingFunctionT}) for two set of constants $c_2$ and $g$ with fixed $p=2$, illustrating the functional form.
(b) and (c) Temperature and field dependencies of the specific heat from the fit to the measured data in Fig~\ref{fig:1}b and c.  
(d) $(C/T)T^{0.54}$ vs. $B/T^{1.08}$ scaling of the measured temperature and magnetic field dependence data from Fig.~\ref{fig:1}. The limiting value for low $B/T^{1+\beta}$ is given by $\gamma_0 T_0^{\alpha_T}$ from Eq.~(\ref{eq:CScaling}).
(e) Intensity plot of chi-square for the optimization of the fitting parameters $\alpha_T$ and $\alpha_B$.
(f) Corresponding intensity plot for the parameter $p$.}
\label{fig:2}
\end{figure*}

Effectively at the quantum critical point, scale invariance will give a zero-field specific heat that increases with decreasing temperature as a power-law \cite{Belitz2005},
\begin{align}\label{Eq:ScalingEquation_SpecificHeat_Tdep}
    \frac{C_\mathrm{el}}{T} \propto T^{- \alpha_T}.
\end{align}
After subtracting the nuclear contribution to $C/T$ in Fig.~\ref{fig:1}b, the remaining $C_\mathrm{el}/T$ closely follows Eq.~(\ref{Eq:ScalingEquation_SpecificHeat_Tdep}).

For large $B$, on the other hand, the specific heat predominantly depends only on $B$ at low $T$, and experimentally follows a similar power-law,
\begin{align}\label{Eq:ScalingEquation_SpecificHeat_Bdep}
    \frac{C_\mathrm{el}}{T} \propto B^{- \alpha_B},
\end{align}
as seen in Fig.~\ref{fig:1}c.
This is consistent with the magnetic energy $\mu_{B} B$ entering as a cutoff parameter in a way similar to the thermal $k_{B}T$. The field and temperature then compete to set the $\xi_{t,\text{eff}}$ and the competition from a temperature-limited to a field-limited quantum critical regime is, at a first approximation, given by
\begin{equation}\label{Eq:Bcutoff}
\eta \frac{\mu_{B}B^\ast}{k_{B}T} = 1  \qquad .
\end{equation}
Here $\eta$ is a constant or weakly temperature-dependent function that should characterize the crossover field $B^\ast(T)$ seen in Fig.~\ref{fig:1}c.
With the limiting cases of Eq.~(\ref{Eq:ScalingEquation_SpecificHeat_Tdep}) and Eq.~(\ref{Eq:ScalingEquation_SpecificHeat_Bdep}), and the crossover field $B^\ast(T)$ giving the thermal-magnetic competition, it is a short step to propose that properties such as magnetic susceptibility and specific heat should display simple scaling relations in the vicinity of the QCP, with $B^\ast(T)$ as a scale for magnetic fields and the corresponding temperature $T^\ast(B)$ as a scale for temperature.

To find a scaling expression for $C_\mathrm{el}/T$ as an explicit function of $T$ and $B$, we use Eq.~(\ref{Eq:ScalingEquation_SpecificHeat_Tdep}) to write 
\begin{equation}\label{eq:CScaling}
\frac{C_\mathrm{el}}{T} = \gamma_0 {\left(\frac{T}{T_0}\right)}^{-\alpha_T} f(x)
\end{equation}
where $\gamma_0$ is the bare electronic Sommerfeld coefficient, $T_0$ is a characteristic temperature scale beyond which scaling is no longer expected, and $f(x)$ should be a function of $B/B^{\ast}(T)$ such that $f(0)=1$.
Electronic specific heat as $C_\mathrm{el}/T$ is proportional to the effective mass. For heavy-fermion compounds, it has been suggested that the electronic effective mass around the quantum critical point can be described using a particular function of $T/T^\ast(B)$ \cite{Shaginyan2010, Shaginyan2013, Shaginyan2016}. Writing this function as a function of $B/B^\ast(T)$ rather than $T/T^\ast(B)$ gives a similar function,
\begin{equation}\label{Eq:ScalingFunctionT}
 f(x \equiv B/B^\ast(T)) = {\frac{1+c_1 x^p}{1+c_2 x^{p+\alpha_B}}} ,
\end{equation}
where $c_1$ and $c_2$ are constants and $p$ is an exponent. We fix the relation between $c_1$ and $c_2$ so that $f(x)$ in Eq.~(\ref{Eq:ScalingFunctionT}) has its maximum at $x=1$, where
\begin{equation}
c_1 = \frac{c_2 (p+\alpha_B)}{p -\alpha_B c_2}
\end{equation}
and $c_1>c_2$. In the limit of $x=0$, this function is constant, while in the limit of large $x$, we would get
\begin{equation}\label{eq:Climit}
\frac{C_\mathrm{el}}{T}\Big|_{\mathrm{high} B} = \gamma_0 \frac{c_1}{c_2}{\left(\frac{T}{T_0}\right)}^{-\alpha_T} {\left(\frac{\eta \mu_B B}{k_BT}\right)}^{-\alpha_B}.
\end{equation}
For Eq.~(\ref{eq:Climit}) to be consistent with Eq.~(\ref{Eq:ScalingEquation_SpecificHeat_Bdep}) and cancel out the temperature dependence, we could either impose that $\alpha_T = \alpha_B$ or we could introduce a weak temperature dependence into $\eta$,
\begin{equation}\label{eq:Eta}
\eta = \eta_0 {\left(\frac{T_0}{T}\right)}^{\beta}
\end{equation}
where $\beta$ should be close to $0$ for Eq.~(\ref{Eq:Bcutoff}) to hold.
Since we expect no temperature dependence of $C/T$ at high fields, as seen in Fig~\ref{fig:1}b, we impose
\begin{equation}\label{eq:beta}
\beta = \frac{\alpha_T-\alpha_B}{\alpha_B}
\end{equation}
This simplifies Eq.~(\ref{eq:Climit}) to
\begin{equation}\label{eq:Climit_2}
\frac{C_\mathrm{el}}{T}\Big|_{\mathrm{high} B} = \gamma_0 \frac{c_1}{c_2} {\left({\frac{\eta_0 \mu_B B}{k_B T_0}}\right)}^{-\alpha_B},
\end{equation}
describing the high-field limit.

We have now arrived at a scaling expression, given by Eq.~(\ref{eq:CScaling}) and Eq.~(\ref{Eq:ScalingFunctionT}) with
\begin{equation}\label{eq:X}
x = \eta_0 \frac{\mu_B B}{k_B T}{\left(\frac{T_0}{T}\right)}^\beta = g {\frac{B}{T^{1+\beta}}}
\end{equation}
where $g = \eta_0 \mu_B T_0^\beta/k_B$, consistent with the limiting behavior of Eq.~(\ref{Eq:ScalingEquation_SpecificHeat_Tdep}) and Eq.~(\ref{Eq:ScalingEquation_SpecificHeat_Bdep}).
We use this scaling function to investigate the scaling behavior of the temperature- and field dependent specific heat of Fig.~\ref{fig:1}. 
Figure~\ref{fig:2}a shows the function $f(x)$ for different sets of parameters $c_2$ and $g$ with fixed $p =2$. The scaling curve is characterized by three quantum critical regions; the temperature-limited (small $x$), crossover ($x=1$), and the field-limited (large $x$) regions.

The data of Fig.~\ref{fig:1}b,c were used to fit an expression for the total specific heat capacity,
\begin{align}\label{Eq:totalspecificheat}
    \frac{C}{T}\Big|_{\mathrm{total}} = \frac{a}{T^3} + \frac{C_\mathrm{el}}{T} + bT^2,
\end{align}
where $C_\mathrm{el}/T$ is given by Eq.~(\ref{eq:CScaling}), the parameter $a$ describes the high-temperature Schottky tail of the nuclear specific heat, and $b$ gives the low-temperature phonon contribution.

The optimum fit was obtained by stepping through $\alpha_T$ and $\alpha_B$, leaving 6 free parameters. The free parameters were optimized by minimizing $\chi^2$, taken as the average mean square of deviations of $T^{\alpha_T}C/T$ for the complete set of data. The obtained fitting parameters are listed in Table~\ref{table1}.

\begin{table}[t!]
\begin{tabular}{ccc}
\hline
Parameter & value & unit\\
\hline
\hline
$a$ & 1.26 & mJ\,K/mol-Yb \\
$b$ & 2.83& mJ/mol-Yb\,K$^4$ \\
$\gamma_0 T_0^{\alpha_T}$ & 240 & (mJ/mol-Yb\,K$^2$)K$^{\alpha_T}$ \\
$\alpha_T$ & 0.54 & \\
$c_1$ & 1.73 & \\
$c_2$ & 1.10 & \\
$p$ & 2.37 & \\
$\alpha_B$ & 0.50 & \\
$g$ & 1.09 & (K/T)K$^\beta$\\
\hline
\end{tabular}
\caption{
Parameters from the fit to Eq.~(\ref{Eq:totalspecificheat}) using Eq.~(\ref{eq:CScaling}) for $C_\mathrm{el}/T$ and Eq.~(\ref{Eq:ScalingFunctionT}) for $f(x)$. The fit does not separate $\gamma_0$ and $T_0$ but gives only the combined $\gamma_0 T_0^{\alpha_T}$.}
\label{table1}
\end{table}

The temperature and field dependence of $C/T$ from the resulting fit with the lowest overall $\chi^2$ is shown in Figure~\ref{fig:2}b,c. As seen by comparing Fig.~\ref{fig:1}b,c and \ref{fig:2}b,c, the general temperature and field dependence are described well by the fit.

Figure~\ref{fig:2}d shows $(C_\mathrm{el}/T)T^{\alpha_T}$ vs. $B/T^{(1+\beta)}$ for all temperature and magnetic field dependence data of Fig.\ref{fig:1} in the region $0.3\,\mathrm{K} < T < 6\,\mathrm{K}$, using values of $\alpha_T = 0.54$ and $\alpha_B = 0.50$ ($\beta = 0.08$) obtained from the best fit as seen in Fig.~\ref{fig:2}e. Note that the scaling analysis when plotted this way depends only on the parameters $\alpha_T$ and $\alpha_B$. The optimized value of $p$ as a function of $\alpha_B$ and $\alpha_T$ is shown in Fig.~\ref{fig:2}f. The best-fit value $p=2.37$ characterizes the crossover region together with $c_2$. From the scaling analysis, we observe that ${\alpha_T \approx \alpha_B \approx 0.5}$, and thus $\beta \approx 0$. This confirms our assumption that magnetic field enters as a cutoff parameter competing with temperature to set the cutoff scale for the quantum criticality.

\begin{figure}[t!!]
\centering
\includegraphics[width=0.9\columnwidth]{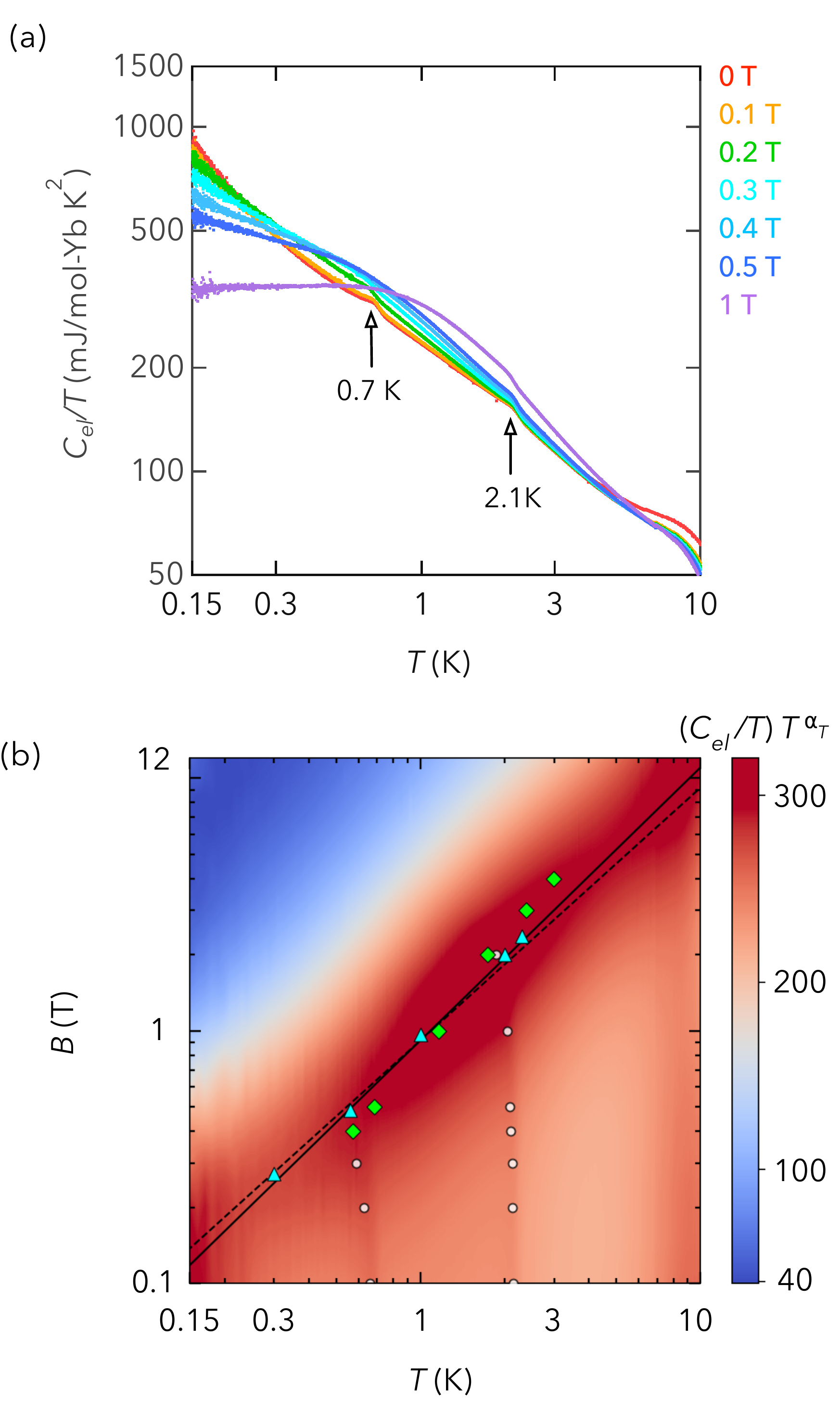} 
\caption{(a) Temperature dependence of electronic specific heat for low fields, highlighting the small transitions/anomalies seen at $\sim0.7$\,K and $\sim2.1$\,K. These anomalies disappear above $0.3$\,T and $2$\,T, respectively, and have a weak field dependence. (b) Intensity plot of $T^{\alpha_T}C_\mathrm{el}/T$ in the $B$\,vs.\,$T$ plane. Solid line indicates the $B^\ast$ from Eq.~(\ref{Eq:Bcutoff}), separating the temperature-limited and field-limited quantum critical regions. Markers represent the experimentally determined crossover fields $B^\ast$ ($\triangle$) and crossover temperatures $T^\ast$ ($\diamondsuit$) for individual measurement curves. Dashed line corresponds to $\mu_B B^\ast / k_B T = 0.62$. The small circles show the low-field transitions/anomalies of panel~a.
	}
\label{fig:3}
\end{figure}

Although hard to spot on the larger scale of Fig.~\ref{fig:1}b, we find that there are two small transitions or anomalies appearing at 0.7\,K and 2.1\,K, for magnetic fields up to about 0.3\,T and 2\,T, respectively. The low-field $C_{el}/T$ between 0.15\,K and 10\,K is shown in Figure~\ref{fig:3}a, illuminating these anomalies. We are confident that these anomalies arise from the bulk of the sample. For a regular non-heavy fermion metal, the associated specific heat signature would be rather significant. However, no anomalies in magnetic susceptibility have been reported at these temperatures \cite{deguchi2012quantum}.

Figure~\ref{fig:3}b shows $T^{\alpha_T}C_{el}/T$ as a function of $T$ and $B$, plotted as an intensity plot. The transitions or features seen in Fig.~\ref{fig:3}a are shown as circular markers. Shifting slightly down in temperature with increasing magnetic field, one could speculate that these transitions may be related to some type of local antiferromagnetic ordering, but we leave the question open for discussion and further analysis.

The maximum in the scaling curve, corresponding to the field-temperature competition crossover, is clearly visible as the dark red region in Fig.~\ref{fig:3}b. The location of $B^\ast(T)$ of Eq.~(\ref{Eq:Bcutoff}), given by $x=1$ and Eq.~(\ref{eq:X}), is marked by the solid line. The weak temperature dependence of $\eta$ is evident if one takes $\beta$ to be zero and plots the corresponding relation with $\eta = \eta_0$, shown as a dashed line in Fig.~\ref{fig:3}b.

\section{Conclusions}
In conclusion, we have measured the low-temperature and magnetic field dependence of specific heat of the Yb-Au-Al quasicrystal. We find that this system is intrinsically quantum critical, with critical exponents $C_{\mathrm{el}}/T\propto T^{-0.54}$ at zero field and $C_{\mathrm{el}}/T\propto B^{-0.50}$ in high magnetic fields. The magnetic field acts as an energy scale competing with temperature to cut off the critical fluctuations. A scaling function is devised that captures the limiting behaviors as well as the $B$-$T$ competition governing the quantum critical scaling of the electronic specific heat.

\section{Acknowledgments}

\begin{acknowledgments}
We thank Arkady Shekhter for illuminating discussions. We thank Michael Sannemo Targama for characterizing the sample composition. This work was supported by the Knut and Alice Wallenberg Foundation (Grant KAW 2018.0019). A.K. and A.R. acknowledge support by the Swedish Research Council, D. Nr. 2021-04360.
\end{acknowledgments}

%

\end{document}